\documentclass[aps,pra,twocolumn, 10pt, showpacs,superscriptaddress,groupedaddress]{revtex4-1}  

\usepackage{graphicx}
\usepackage{epstopdf}
\usepackage{bbold}
\usepackage{comment}
\usepackage{color}
\usepackage{soul}
\usepackage[abs]{overpic}
\usepackage{amsmath,amsfonts,amssymb}
\usepackage{lipsum}
\usepackage{hyperref}
\usepackage{setspace}
\graphicspath{{Figures/}}
\begin{document}
%
\author{Oz Livneh}
\affiliation{Department of Physics of Complex Systems, Weizmann Institute of Science, Rehovot 7610001 , Israel }
\author{Gadi Afek}
\affiliation{Department of Physics of Complex Systems, Weizmann Institute of Science, Rehovot 7610001 , Israel }
\author{Nir Davidson}
\affiliation{Department of Physics of Complex Systems, Weizmann Institute of Science, Rehovot 7610001 , Israel }
\title[]{Producing an Efficient, Collimated and Thin Annular Beam with a Binary Axicon}
\begin{abstract}
We propose and demonstrate a method to produce a thin and highly collimated annular beam that propagates similarly to an ideal thin Gaussian ring beam, maintaining its excellent propagation properties. Our optical configuration is composed of a binary axicon - a circular binary phase grating, and a lens, making it robust and well suited for high-power lasers. It has a near-perfect circular profile with a dark center, and its large radius to waist ratio is achieved with high conversion efficiency. The measured profile and propagation are in excellent agreement with a numerical Fourier simulation we perform.
\end{abstract}
\maketitle
Annular beams, characterized by a hollow ring cross section, are useful for a variety of applications, such as optical dipole traps for ultra-cold atoms~\cite{Nir_2000,Nir_2002,Bigelow_2008,Hadzibabic_2013,Gaunt2015,Zwierlein_2017}, hollow optical tweezers for dielectric particles~\cite{Padgett_1998Tweez,Padgett_2000}, imaging and super-resolution microscopy~\cite{Hell_1994,Novotny_2000}, long-ranged atmospheric optical communication~\cite{Phillips_2005, Baykal_2006} and material processing~\cite{Kar_2005,Kar_2006}.

There are many techniques to produce annular beams. However, it is challenging to produce an ideal Gaussian ring beam that is minimally diffracting. Usually there is a compromise between ring parameters (thinness, darkness at the center), beam propagation (minimal diffraction, collimation, shape invariance), complexity, power handling and efficiency. Inter-cavity techniques for lasing in annular modes~\cite{Hasman_2001,Asher_2005} may have a Gaussian-like profile with minimal and symmetric diffraction, at the expense of low gain and output versatility. Extra-cavity techniques to transform from Gaussian to annular beams can produce thin and dark beams with asymmetric or strong diffraction~\cite{Nir_2000, Nir_2002}, or efficient Gaussian-like annular beams with limited geometry~\cite{Grimm_1998,Padgett_1998LG,Abraham_2002,Hennequin_2002,Tino_2003,Lin_2005}. 

Widely used Laguerre Gaussian (LG) beams, $LG_p^l$ of radial index $p=0$ and azimuthal index $l$, have a Gaussian-like cylindrical profile with a radius to waist ratio of $\sqrt{l/2}$~\cite{Dholakia_2006}. The highest efficiency of converting a Gaussian beam into $LG_0^1$ by typical extra-cavity techniques is $\approx 80\%$~\cite{Abraham_2002}, but the overlap between a Gaussian beam and $LG_0^l$ decays significantly with $l$. For a thin LG beam with radius to waist ratio of $\approx 10$ it is practically zero~\cite{Padgett_1998LG}.

Many techniques to create annular beams, mostly extra-cavity, involve an axicon~\cite{Grimm_1998,Hennequin_2002,Tino_2003,Lin_2005,Dickey2000,Aieta2012,Lin2014,Lalithambigai2012}, a refractive conical optical element which inflicts a linear radial phase to the refracted light. An axicon ideally transforms a Gaussian beam into a Bessel beam~\cite{Arrizon2009,McLaren2012,McLaren2013,Arrizon2014}, which opens to a ring with a cylindrical profile of half a 1D Gaussian beam. The main drawback of the axicon is the inherent imperfection of its tip area, which cannot be infinitely polished to a point. The deviation of the output beam from the expected output, which is itself a non-ideal half Gaussian, is stronger for thinner rings requiring a smaller input waist that increases the overlap with the imperfect tip.

In this Letter we discuss the propagation of Gaussian ring beams. We propose and demonstrate a novel method to produce a thin and highly collimated annular beam that propagates as, and even outperforms in a way, an ideal Gaussian ring beam, with a simple optical configuration composed of a diffractive binary axicon, a circular binary phase element, and a lens. Albeit less versatile, the binary axicon is much cheaper and (being a passive element) more robust than spatial light modulators~\cite{Padgett_1998LG,Dholakia_2006,Hadzibabic_2013}. Furthermore, it is well suited for high-power lasers, displaying damage thresholds higher than those of the typical spatial light modulator by many orders of magnitude. In contrast to LG beams, high radius to waist ratio of $\approx 10$ is achieved with high efficiency. As opposed to the output of a refractive axicon, the thin ring created by our setup has a Gaussian profile. It is constructed from combined diffraction orders of the binary axicon, thus ensuring $>80\%$ efficiency even for a binary element. Its Rayleigh range is longer by $\approx50\%$ than that of an ideal Gaussian ring beam. The measured profile and propagation are in excellent agreement with our numerical Fourier simulation. 

\begin{figure} 
\centering
\begin{overpic}
[width=\linewidth]{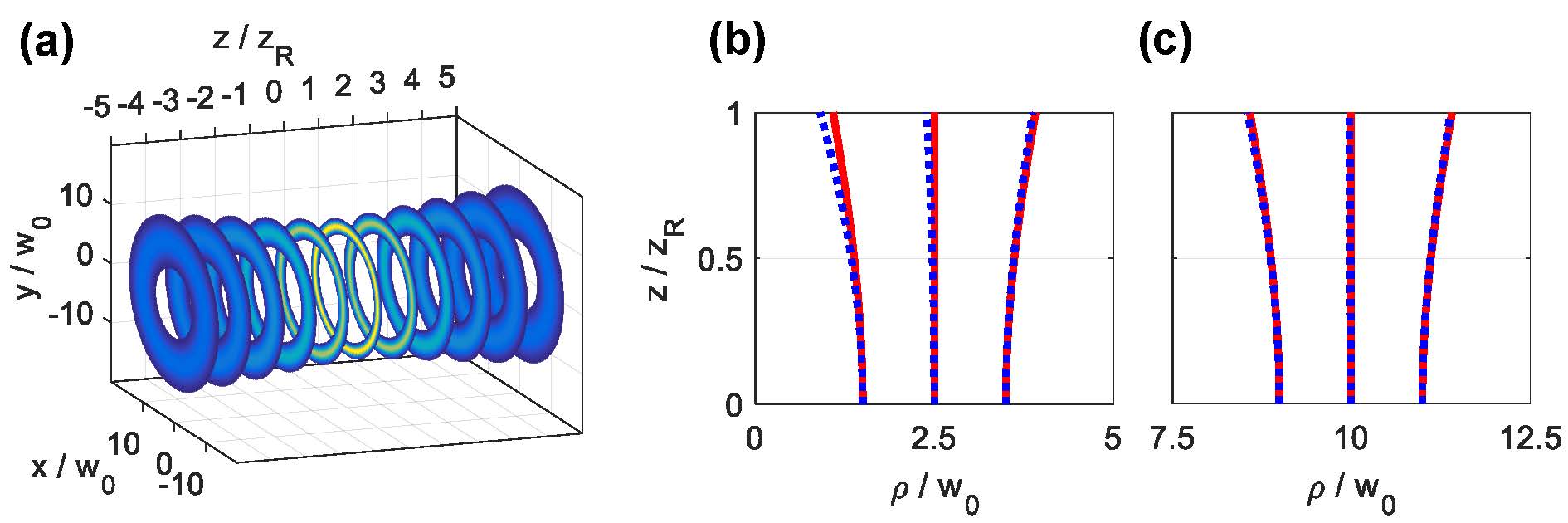}
\end{overpic}
\caption{{Propagation of Gaussian ring beams. }{\bf(a)} Propagation of an ideal Gaussian ring beam, in which the walls diffract as a 1D flat-phase Gaussian beam, with radius to waist ratio, $hollowness\equiv R/w_0=10$. Intensity is described by the color scale (a.u.). {\bf(b, c)} Line of maximal intensity, and the waist lines from both sides - for the simulated propagation (dotted blue), and an ideal Gaussian ring beam with the same parameters (solid red), for {\bf(b)} $hollowness=2.5$, and {\bf(c)} $hollowness=10$. While the vertical axis is the same for both, the horizontal axis limits are different to allow comparison on a similar scale.}
\label{fig:fig1}
\end{figure}

A Gaussian ring beam has, at the focal plane, a ring profile of a flat-phase 1D Gaussian beam. To be ideal, it is required to perfectly maintain the Gaussian ring profile along propagation, illustrated in Fig.~\ref{fig:fig1}(a). This requires the radius to waist ratio $R/w_0$ (hereon defined as $hollowness$) to be large enough such that the ring curvature is negligible. Fig.~\ref{fig:fig1}(b-c) presents numerical simulations, showing how a beam that starts as a Gaussian ring with $hollowness=2.5$ propagates with only a small deviation from an ideal profile, whereas the deviation of a beam that starts with $hollowness=10$ is already completely negligible.

\begin{figure} 
\centering
\begin{overpic}
[width=\linewidth]{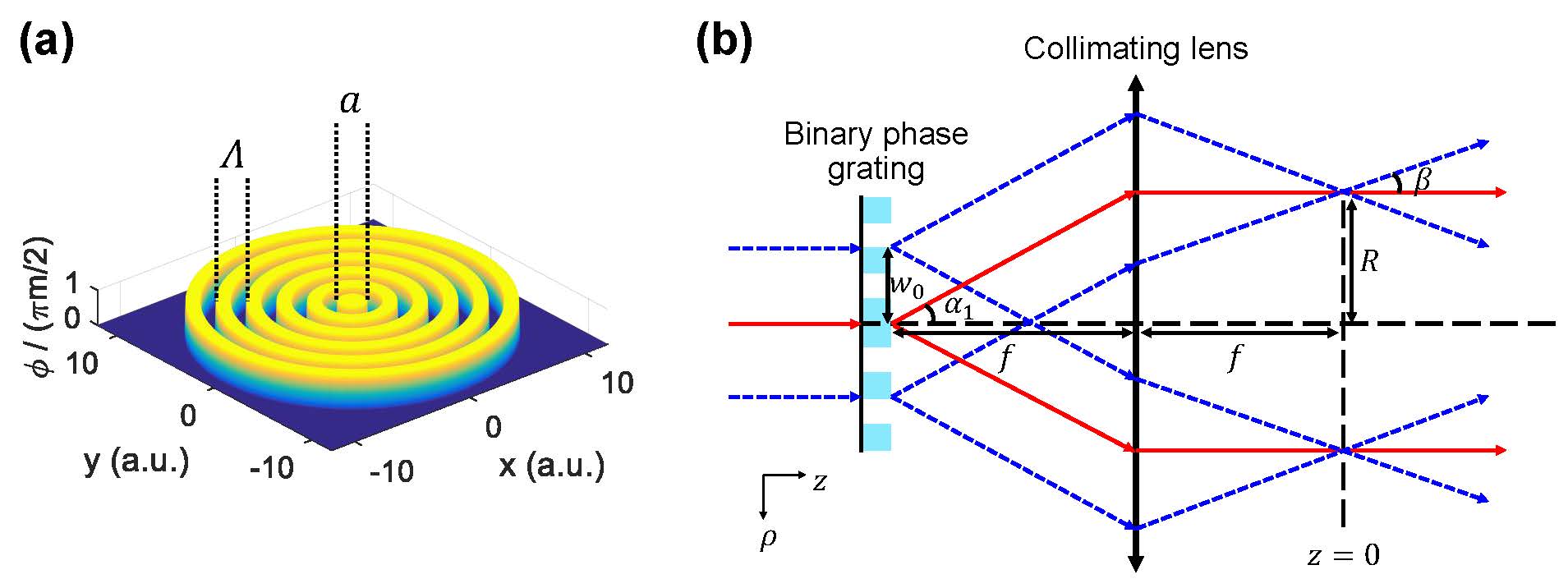}
\end{overpic}
\caption{{\bf(a)} The phase $\phi (x,y)$ inflicted by a binary axicon, a circular phase grating, normalized by $m$, the modulation depth. {\bf(b)} Ray diagram of our optical setup (exaggerated angles and proportions). The input beam diffracts from the binary axicon, placed at the back focal plane of a $f>0$ lens, creating a ring with head angle $\alpha_1$ (only 1$^\text{st}$ diffraction order is illustrated) and theoretical efficiency $\eta_1\approx81\%$. The exterior rays (dashed blue) intersect with the central rays (solid red) at the front focal plane, $z=0$, producing a collimated ring.}
\label{fig:fig2}
\end{figure}

The key element in our scheme is a binary axicon, a circular phase grating, that imprints a phase to light that passes through as described in Fig.~\ref{fig:fig2}(a): $\phi(r)= \frac{m}{2}\times\begin{cases} \pi, & [r/(\Lambda/2)]\in\text{odd} \\ 0, & [r/(\Lambda/2)]\in\text{even} \end{cases}$, where $m$ is the modulation depth, $[...]\equiv floor(...)$, $r=\sqrt{x^2+y^2}$ is the radial coordinate and $\Lambda$ is the grating period. The surface depth is $d (r)=\phi(r)$⁄$k(n-1)$, where $k= 2\pi/\lambda$, $\lambda=780~$nm is the laser wavelength and $n$ is the refractive index. In order to eliminate the zeroth order of the produced ring, the diameter of the innermost grating circle is chosen to be $a=\Lambda$.

The far-field of a radially periodic binary axicon with a plane wave input is expressed analytically as a sum over discrete diffraction orders. First the radial periodicity is exploited to span the input field as a radial trigonometric Fourier series, and only then a cylindrical Fourier transform (Hankel transform) is performed to obtain the scaling and efficiency of the $n^\text{th}$ diffraction order:
\begin{subequations}
\begin{align}
&\tan(\alpha_n)=\frac{\lambda}{\Lambda}n\label{eq:scaling}\\
&\eta_n=\begin{cases} \frac{1}{4}\left\vert1+e^{im\pi/2}\right\vert^2, & n=0 \\ 2\times\left\vert\frac{1}{2\pi n}\left[1-(-1)^n\right]\left(1-e^{im\pi/2}\right)\right\vert^2, & n>0 \end{cases}\label{eq:efficiency}
\end{align}\label{eq:order_and_efficiency}
\end{subequations}

Each $n^\text{th}$ diffraction order is a ring with an opening angle $\alpha_n$, [illustrated for the $1^\text{st}$ order ring in Fig.~\ref{fig:fig2}(b)], and diffraction efficiency $\eta_n$. Generally, even orders vanish because of symmetric interference, except for the $0^\text{th}$ order that vanishes only for a proper choice of $m$. The $n^\text{th}$ ring is constructed from combined $+n$ and $-n$ local diffraction orders, resulting in $\eta_n$ being exactly twice that of a 1D binary phase grating. For an efficient annular beam, the same value of $m=2$ that gives the maximal grating phase difference, $\Delta\phi=\pi$, is required both to cancel the $0^\text{th}$ order and to maximize the $1^\text{st}$ order efficiency in~\eqref{eq:efficiency}. This yields $\eta_1=8/\pi^2\approx81\%$, indeed twice the maximal $1^\text{st}$ order efficiency of a 1D binary phase grating, which is $\approx40.5\%$~\cite{Goodman_FourierOptics}.

To collimate the $1^\text{st}$ order ring, the binary axicon is placed at the back focal plane of a $f>0$ lens, as illustrated in Fig.~\ref{fig:fig2}(b). The input is a Gaussian beam with waist $w_0$. The exterior rays (dashed blue) intersect with the central rays (solid red) at the front focal plane, producing a collimated ring. The waist $w$ at the focal plane can be approximated as the waist $w_f$ of the input Gaussian beam after being focused by the lens:
\begin{equation}
    w\approx w_f=\lambda f/(\pi w_0)
    \label{eq:wf}
\end{equation}
Combining~\eqref{eq:scaling} and~\eqref{eq:wf} and defining $R_1$ as the radius of the first-order ring yields:
\begin{equation}
    hollowness=\frac{R_1}{w}\approx\frac{\pi w_0}{\Lambda}
    \label{eq:hollowness}
\end{equation}

The element used is a surface-relief binary axicon (Holo-Or, custom-made), with $\Lambda\approx402~\mu$m and transmission efficiency of $\approx100\%$. The remaining configuration parameters are $w_0$ and $f$. The input originates from a TEM$_{00}$ single-mode fiber of $w_0=1.1$~mm constraining a pure mode, propagates through the binary axicon and is then collimated by a $f=75$~mm lens.

\begin{figure} 
\centering
\begin{overpic}
[width=\linewidth]{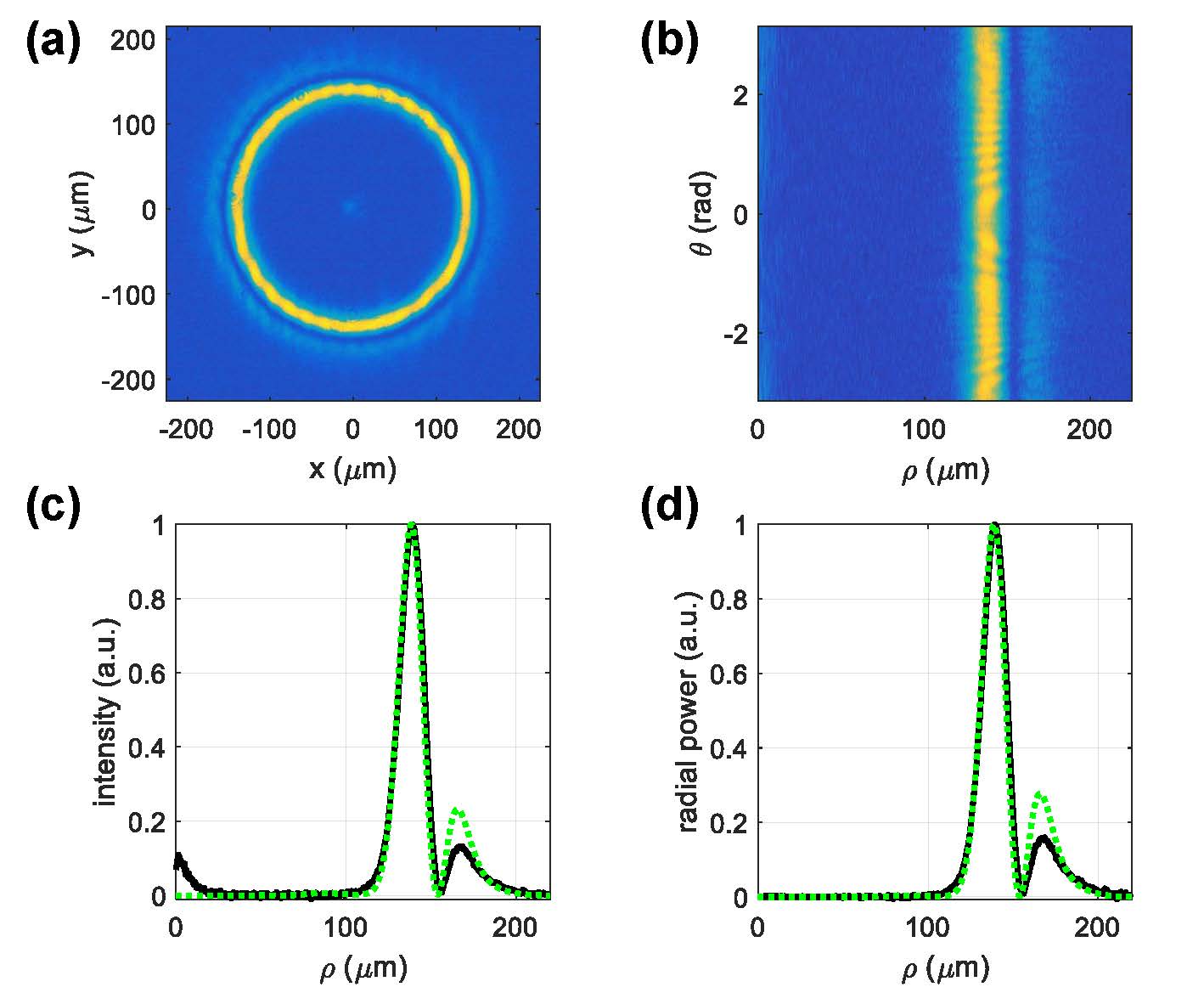}
\end{overpic}
\caption{{Profile analysis at the focal plane.} {\bf(a)} Measured $xy$ intensity. {\bf(b)} $r\theta$ intensity, achieved by a polar transformation of (a). {\bf(c)} Radial intensity (solid black), calculated by integrating over $\theta$ in (b), from which $hollowness\approx10$ is extracted, plotted together with the simulated profile (dotted green). {\bf(d)} Radial power (a.u. of power/length), obtained by multiplying (c) by $r$. The area under this graph is the power.}
\label{fig:fig3}
\end{figure}

Fig.~\ref{fig:fig3}(a) presents the measured intensity at the focal plane. It is a near-perfect circular ring surrounded by a weaker ring, translated into straight lines in the polar representation of Fig.~\ref{fig:fig3}(b). The deviation of the radius line from its mean is $\approx0.1\%$, and there exists a residual DC component of $\sim10\%$ of the maximal intensity. This can arise from a small residual wavelength mismatch\footnote{In order to compensate for a small wavelength mismatch, we tilted the grating relative to the incident beam, thus effectively decreasing the acquired phase, $\phi_\text{tilt}=\phi\cos\theta$. The tilt angle is $\theta=\cos^{-1} (780/808)\approx0.264$~rad, relatively small.}, and can be totally removed using a spatial filter such as a blackened sphere~\cite{Nir_2002}. Fig.~\ref{fig:fig3}(c-d) present the measured radial profile together with a corresponding numerical Fourier simulation, calculated by propagating a flat-phase Gaussian input beam through the binary axicon and obtaining its far-field. There is an excellent agreement between the simulation and the measured main lobe of the $1^\text{st}$ order ring.

We also observe a weak side lobe, both in our experimental and numerical results. It is part of the $1^\text{st}$ diffraction order, and is related to the singularity at the origin, characterized by the parameter $a$ defined in Fig.~\ref{fig:fig2}(a). We find that only for $a=\Lambda$ and $m=2$ the $0^\text{th}$ order vanishes.

The main lobe in Fig.~\ref{fig:fig3}(c) is well fitted by a Gaussian to extract the ring parameters. Its peak radius is $138.4\pm0.1~\mu$m. To compare with the $R_1\approx145~\mu$m obtained from~\eqref{eq:scaling} that corresponds to the complete diffraction order and not only the main lobe, we calculate the "center of mass" of the double feature to obtain a radius of $142.3~\mu$m, in full agreement. The waist is $w=14.1\pm0.1~\mu$m, close to the approximated $w_f\approx17~\mu$m obtained from~\eqref{eq:wf}. This yields $hollowness\approx9.8$, close to the approximated value $\approx8.6$ obtained from~\eqref{eq:hollowness}.

The power in cylindrical coordinates is given by the area below $rI(r)$ in Fig.~\ref{fig:fig3}(d). The center of the beam is extremely dark except for a small $0^\text{th}$ order peak, containing $\approx0.7\%$ of the main ring energy. The $1^\text{st}$ order diffraction efficiency is measured with a power-meter, as the power when an iris is properly closed around the $1^\text{st}$ order ring, out of the total power. This yielded $\eta_1=86\pm3\%$, in reasonable agreement with the expected $\eta_1\approx81\%$ from~\eqref{eq:efficiency}.

\begin{figure} 
\centering
\begin{overpic}
[width=\linewidth]{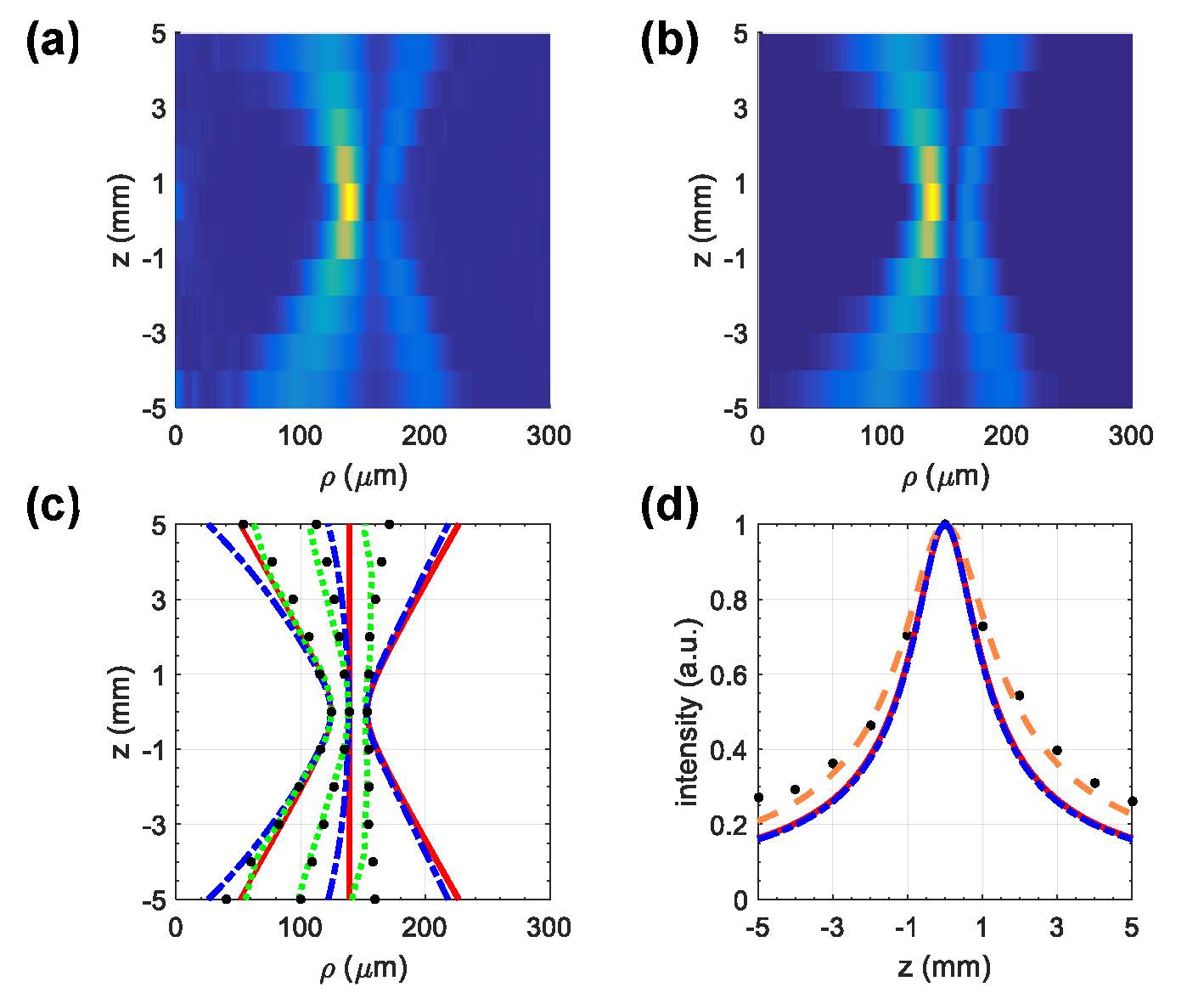}
\end{overpic}
\caption{{Propagation analysis.} {\bf(a)} Measured and {\bf(b)} simulated intensity profile (color scale, a.u.). The horizontal and vertical axes are of different scales. {\bf(c)} Comparison of the main lobe line of maximal intensity, and waist lines from both sides, for the measurement (black points), simulation (dotted green), a numerically propagated Gaussian ring of the same parameters (dash-dotted blue), and an ideal Gaussian ring beam (solid red). {\bf(d)} Longitudinal decay of the maximal intensity (black points), closely fitted by forward and backward longitudinal decays of a 1D Gaussian beam (dashed orange), from which we extract the Rayleigh range $1.2\pm0.06$~mm that is longer by 50\% than that of the geometrically fitted ideal Gaussian ring beam (solid red) or that of the propagated Gaussian ring (dot-dashed blue) from (c).}
\label{fig:fig4}
\end{figure}

Extending the measurement to different $z$ planes enables a construction of the propagation intensity profile, presented in Fig.~\ref{fig:fig4}(a) alongside its Fourier simulation in Fig.~\ref{fig:fig4}(b). Excellent agreement is observed in Fig.~\ref{fig:fig4}(c) between the measured (black points) and simulated (dotted green) main lobe line of maximal intensity, and waist lines from both sides. We compare with a numerically simulated Gaussian ring beam of the same parameters (dot-dashed blue), and with an ideal Gaussian ring beam (solid red). The measured and simulated radius lines bend more than that of the propagated Gaussian ring. The forward ($z>0$) and backward ($z<0$) bending angles of the measured radius line are $\beta_{z>0}^\text{bend}=5\pm2$~mrad, $\beta_{z<0}^\text{bend}=8.8\pm0.4$~mrad, characterizing a highly collimated beam. The measured and simulated inward waist lines closely follow those of an ideal Gaussian, unlike the propagated Gaussian ring that diffracts stronger. Even more interesting, the measured and simulated outward waist lines are almost straight.

Fig.~\ref{fig:fig4}(d) presents the longitudinal intensity decay along the radius line (black points), well fitted by the longitudinal decay of a 1D Gaussian beam $\propto1/w(z)\propto1/\sqrt{1+(z/z_R^\pm)^2}$ (dashed orange), from which we extract the forward and backward Rayleigh ranges, $z_R^+=1.2\pm0.04$~mm and $z_R^-=1.1\pm0.05$~mm. Their mean is longer by $\approx50\%$ than that of the compared ideal Gaussian ring beam [solid red in Fig.~\ref{fig:fig4}(c-d)], $z_R=0.80\pm0.01$~mm, due to the fact that the ring main lobe opens inwards similar to a Gaussian, but almost does not open outwards.

In summary, we discussed the propagation of Gaussian ring beams, and demonstrated a novel technique to produce a thin and highly collimated annular beam that propagates as, and even outperforms, an ideal Gaussian ring beam. Our simple optical configuration is robust and well suited for high-power lasers. High radius to waist ratio of $hollowness\approx 10$ is achieved with efficiency of $86\pm3\%$, twice that of a 1D binary phase grating. Our beam is dark at the center, and the ring has a Gaussian profile. 

The measured profile and propagation are in excellent agreement with our numerical Fourier simulation. The main ring lobe opens inwards similar to an ideal 1D Gaussian beam, but almost does not open outwards due to "support" from the weak side lobe similar to Bessel or Airy beams. Thus, its depth of focus is longer by $\approx50\%$ than that of the fitted ideal Gaussian ring beam. Different values of $hollowness$ can be readily obtained without replacing the binary axicon, by changing the waist of the input beam. It can further be used in material processing or as hollow optical tweezers for dielectric particles.

\begin{acknowledgements}
The authors would like to thank Lior Shachaf, Asher Frisem and Dan Oron for discussions.
\end{acknowledgements}


\bibliographystyle{apsrev4-1}
\bibliography{Bibliography}


\end{document}